\documentclass[10pt,aps,prl,twocolumn,nofootinbib,floatfix]{revtex4-2}
\usepackage[T1]{fontenc}
\usepackage{amsmath,amssymb,bm,slashed,graphicx}
\graphicspath{{figures/}{paper/figures/}}

\begin{document}

\title{High-Energy Asymptotics of Quark--Gluon Correlations}
\author{Jian Zhou}
\email{jzhou@sdu.edu.cn}
\affiliation{Key Laboratory of Particle Physics and Particle Irradiation (MOE),
Institute of Frontier and Interdisciplinary Science, Shandong University,
Qingdao, Shandong 266237, China}
\affiliation{Southern Center for Nuclear-Science Theory (SCNT), Institute of
Modern Physics, Chinese Academy of Sciences, Huizhou, Guangdong 516000, China}
\date{\today}

\begin{abstract}
The internal structure of the proton is encoded not only in single-parton densities, but also in the quantum interference between distinct partonic configurations. We study the high-energy behavior of this interference by focusing on flavor-nonsinglet quark--gluon correlations relevant to transverse-spin phenomena. By establishing recursive factorization at double-logarithmic accuracy in the presence of a coherent-gluon insertion, we derive infrared evolution equations for the Qiu--Sterman function and its chiral-odd Boer--Mulders counterpart. We find that these twist-3 functions share the leading small-$x$ intercepts of their respective twist-2 counterparts, but acquire an additional $1/\ln(1/x)$ suppression due to the color structure of the coherent-gluon insertion. These results may guide small-$x$ extrapolations of quark--gluon correlations at the Electron--Ion Collider (EIC).
\end{abstract}

\maketitle

\emph{Introduction.}---Understanding the proton as a quantum many-body system requires going beyond parton probabilities to quantum interference. Quark--gluon correlations encode such interference between distinct partonic configurations. An important example is the twist-three Qiu--Sterman (QS) function~\cite{Qiu:1991pp}, whose diagonal component $T_F^q(x,x)$ probes light-front zero modes~\cite{Ji:2020baz}. This diagonal component is related to the first transverse moment of the Sivers distribution~\cite{Boer:2003cm}, linking zero modes to T-odd spin--momentum correlations. Efforts to determine the QS function span measurements of spin asymmetries \cite{COMPASS:2012dmt,COMPASS:2017jbv}, lattice calculations \cite{Musch:2011er,Yoon:2017qzo}, theoretical analyses~\cite{Ji:2006vf,Zhou:2009jm,Zhou:2010ui,Kang:2011jw,Schafer:2014zea,Zhou:2015ima,Zhou:2017mpw,Scimemi:2019gge,Ji:2020jeb,delRio:2024vvq,Chen:2016dnp,Chen:2017lvx,Liu:2018trl,Liu:2020dct,Liu:2020jjv,Rein:2025pwu,Rein:2025qhe}, and global fits~\cite{Echevarria:2014xaa,Cammarota:2020qcw,Echevarria:2020hpy,Bury:2020vhj,Bury:2021sue,Bacchetta:2020gko,Gamberg:2022kdb,Zeng:2024gun,Vladimirov:2025qrh,Portela:2026wwn}. The Electron--Ion Collider (EIC) will extend transverse-spin measurements into the small-$x$ regime~\cite{Accardi:2012qut,AbdulKhalek:2021gbh}. A pressing theoretical question is therefore how multiparton interference behaves at small $x$.

To isolate the high-energy dynamics of quark--gluon interference, we study the flavor-nonsinglet combination $T_{F,\mathrm{NS}}=T_F^u+T_F^{\bar u}-T_F^d-T_F^{\bar d}$, which cancels flavor-independent mixing with the $C$-even and $C$-odd twist-3 tri-gluon correlators~\cite{Kang:2008ey,Braun:2009mi}. The latter exhibit distinct small-$x$ dynamics~\cite{Schafer:2013opa}: the $C$-odd correlator is associated with the three-gluon limit of the spin-dependent odderon~\cite{Zhou:2013gsa} and the dipole-type gluon Sivers distribution, whereas the $C$-even one is linked to the suppressed Weizs\"acker--Williams-type gluon Sivers distribution~\cite{Boer:2015vso,Boer:2015pni,Szymanowski:2016mbq,Dong:2018wsp,Yao:2018vcg,Boussarie:2019vmk,Boer:2022njw,Mantysaari:2025mht,Schafer:2013wca}.

In the dilute regime, Balitsky--Fadin--Kuraev--Lipatov (BFKL) evolution resums the leading small-$x$ logarithms in gluon distributions~\cite{Kuraev:1977fs,Balitsky:1978ic}.
In contrast, flavor-nonsinglet quark distributions have no leading $1/x$ contribution, and their small-$x$ behavior is instead governed by double logarithms $(\alpha_s\ln^2(1/x))^n$.
These double logarithms can be resummed using the infrared evolution equation (IREE) method~\cite{Kirschner:1982qf,Kirschner:1983di,Ermolaev:1995fx,Bartels:1995iu}.
Previous IREE studies addressed the unpolarized nonsinglet PDF, the polarized structure functions $g_1$ and $g_2$, and the Wandzura--Wilczek part of parton orbital angular momentum~\cite{Ermolaev:1995fx,Bartels:1995iu,Bartels:1999zs,Bartels:2001ve,Boussarie:2019icw}.
In parallel, Kovchegov and collaborators have developed a Wilson-line framework for spin dynamics at small $x$, including corrections beyond the eikonal approximation.
Within this approach, helicity~\cite{Adamiak:2023okq,Dumitru:2024pcv,Adamiak:2025mdy}, transversity, orbital angular momentum, and quark TMDs~\cite{Kovchegov:2015pbl,Kovchegov:2018znm,Kovchegov:2018zeq,Cougoulic:2022gbk,Borden:2023ugd,Kovchegov:2023yzd,Borden:2024bxa,Borden:2025ehe,Kovchegov:2022kyy,Kovchegov:2021iyc,Adamiak:2024khm,Santiago:2023rfl,Caucal:2026yzl} have been studied.

In this Letter, we extend infrared evolution to the QS function. Compared with an ordinary quark density, the QS correlation requires an additional coherent-gluon insertion in the hard scattering. In covariant gauge, we extract the twist-three contribution by differentiating the hard amplitude with respect to the coherent gluon's transverse momentum. We show that this derivative preserves the recursive factorization of ladder and nonladder radiation at double-logarithmic accuracy. The color generator associated with the coherent gluon, together with the quark--antiquark crossing relation, selects a negative-signature color-octet amplitude, in contrast to the color-singlet channel governing the corresponding unpolarized flavor-nonsinglet quark distribution. Solving the resulting coupled IREEs yields the asymptotic small-$x$ behavior
\begingroup
\setlength{\abovedisplayskip}{5pt}
\setlength{\belowdisplayskip}{5pt}
\setlength{\abovedisplayshortskip}{5pt}
\setlength{\belowdisplayshortskip}{5pt}
\begin{equation}
 T_{F,\mathrm{NS}}(x,x;Q^2)\propto
 \frac{x^{-\sqrt{8\alpha_s/(3\pi)}}}{[\ln(1/x)]^{5/2}}
 \label{eq:qs-asymptotics-intro}
\end{equation}
\endgroup
at fixed $Q^2$. Thus, $T_{F,\mathrm{NS}}$ shares the same leading intercept as its unpolarized twist-2 counterpart, but acquires an extra inverse logarithm in $x$. We further apply our approach to the chiral-odd quark--gluon correlator associated with the Boer--Mulders function and find the same additional inverse-logarithmic suppression relative to the transversity distribution.

\emph{Quark--gluon--quark correlation function.}---The twist-three Qiu--Sterman function is defined as
\begin{equation}
 \begin{aligned}
 &\epsilon_T^{ij}S_{Tj}\,T_F^q(x_1,x_2;\mu^2)\\
 &\quad=\int\frac{\mathrm d\xi^-\,\mathrm d\eta^-}{4\pi}\,
 e^{ix_1P^+\xi^-+i(x_2-x_1)P^+\eta^-}\\
 &\qquad\times\bigl\langle P,S_T\big|
 \bar q(0)\gamma^+gF^{+i}(\eta^-)q(\xi^-)
 \big|P,S_T\bigr\rangle .
 \end{aligned}
 \label{eq:qs-operator-definition}
\end{equation}
Here $P$ and $S_T$ are the nucleon momentum and transverse spin vector, respectively, and $\mu$ is the renormalization scale. We set $\xi^+=\eta^+=0$ and $\bm\xi_T=\bm\eta_T=0$. The variables $x_1$ and $x_2$ are the quark longitudinal momentum fractions. The straight lightlike Wilson lines ensuring gauge invariance are left implicit.

The diagonal matrix element $T_F^q(x,x)$ is related to the first $k_T$-moment of the bare, unsubtracted transverse-momentum-dependent (TMD) Sivers function $f_{1T,\mathrm{unsub}}^{\perp q[\mathrm{DY}]}$, defined with a past-pointing Drell--Yan (DY) staple gauge link. In our sign convention, the bare-operator relation is written as \cite{Boer:2003cm}
\begin{equation}
 T_F^q(x,x)
 =\frac{1}{M}\int\mathrm d^2\bm k_T\,\bm k_T^2\,
 f_{1T,\mathrm{unsub}}^{\perp q[\mathrm{DY}]}(x,\bm k_T^2).
 \label{eq:sivers-moment}
\end{equation}
Here $M$ is the nucleon mass and $\bm k_T$ is the measured quark transverse momentum. We write $k_T^2\equiv\bm k_T^2$, with analogous notation for other transverse momenta. We use Eq.~\eqref{eq:sivers-moment} to calculate the QS function from the Sivers function, so that we can treat the gauge-link contributions explicitly in cut diagrams. In this representation, we include gluon attachments to the past-pointing Wilson lines together with those to the hard part~\cite{Ji:2002aa,Belitsky:2002sm}. After the weighted transverse-momentum integration, the light-cone divergences cancel exactly between the real and virtual contributions.

In covariant gauge, the $-\partial_\perp^i A^+$ piece of $F^{+i}$ is extracted by differentiating the hard amplitude with respect to the coherent-gluon transverse momentum $\bm p_T$. To obtain the T-odd contribution, we couple the coherent gluon to the past-pointing Wilson lines and include the conjugate mirror diagrams \cite{Ji:2006vf,Eguchi:2006mc,Zhou:2009jm}. The required imaginary part is generated by soft-gluon poles (SGPs) and hard-gluon poles (HGPs). To identify the logarithmically enhanced HGP contribution at small $x$, we consider the region of strong rapidity ordering. In this region, nearly all of the parent quark's longitudinal momentum is carried by the emitted gluon. The HGP condition then forces the coherent gluon to carry a comparably large longitudinal momentum fraction. We therefore consider ladder and nonladder diagrams with quark--gluon exchange in the $t$ channel. For fixed external transverse momentum, the region of much larger internal transverse momenta is suppressed by the real-gluon-emission kernel in this channel~\cite{Kirschner:1994rq,Kirschner:2005rj}. The transverse-momentum integral over this region supplies no second logarithm, so this gluon-emission contribution is subleading at double-logarithmic accuracy (DLA).

\begin{figure}[!htbp]
 \centering
 \includegraphics[width=\columnwidth]{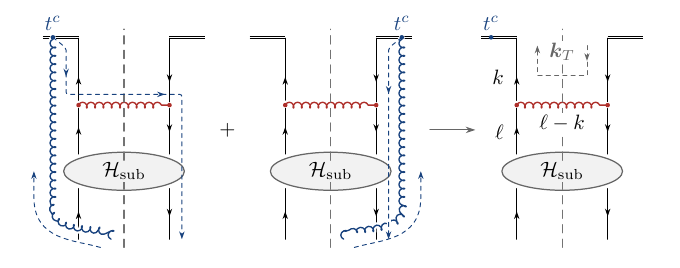}
 \caption{Mirror diagrams contributing to the SGP term. Dashed blue and grey arrows indicate the transverse-momentum flow associated with $\bm p_T$ and $\bm k_T$, respectively. The observed and parent quark momenta are $k$ and $\ell$.}
 \label{fig:sgp-mirror-pair}
\end{figure}

\emph{High-energy evolution of quark--gluon correlations.}---We now show that the $\bm p_T$ derivative preserves the recursive factorization of the small-$x$ evolution. We begin with the ladder diagrams in Fig.~\ref{fig:sgp-mirror-pair}. Using the SGP master formula~\cite{Eguchi:2006mc,Koike:2006qv}, we express their spin-dependent contribution in terms of a transverse derivative of the corresponding unpolarized hard part $\mathcal H$, evaluated without the coherent-gluon attachment.

To identify the leading double logarithms, we consider a parent quark $\ell$ with longitudinal momentum fraction $\beta\gg x$ and transverse momentum $\bm\kappa_T\equiv\bm\ell_T$. In the logarithmically enhanced region, its light-front lifetime, proportional to $\beta/\kappa_T^2$, is much longer than that of the observed quark, proportional to $x/k_T^2$~\cite{Caucal:2022ulg}. We therefore integrate over transverse momenta up to $\kappa_T^2\sim(\beta/x)k_T^2$, rather than imposing strong ordering in transverse momentum. Together with the longitudinal integration over $\mathrm d\beta/\beta$, this region supplies the double logarithms resummed by infrared evolution~\cite{Ermolaev:1995fx,Boussarie:2019icw}. At fixed $\beta$, we can thus write the ladder contribution schematically as
\begin{equation}
 \mathcal H(k_T^2)
 =\frac{1}{k_T^2}\int_{\mu^2}^{(\beta/x)k_T^2}
   \frac{\mathrm d\kappa_T^2}{\kappa_T^2}\,
   \mathcal H_{\rm sub}(\beta,\kappa_T^2).
 \label{eq:ladder-hard-part}
\end{equation}
We have extracted the $1/k_T^2$ dependence of the uppermost rung and collected the remaining ladder in $\mathcal H_{\rm sub}$.

For the SGP mirror pair, the transverse recoil enters the hard part through $\bm k_T-\bm p_T$. Differentiating with respect to $p_T^i$ and then setting $\bm p_T=0$, we obtain $-\partial\mathcal H/\partial k_T^i$. The Sivers projection contracts this derivative with $k_T^i$, while azimuthal averaging introduces a factor of $1/2$. We thus obtain
\begin{equation}
 \begin{aligned}
 -\frac{k_T^i}{2}\frac{\partial\mathcal H}{\partial k_T^i}
 &=-k_T^2\frac{\partial\mathcal H}{\partial k_T^2}\\
 &=\mathcal H-\frac{\mathcal H_{\rm sub}(\beta,(\beta/x)k_T^2)}{k_T^2}
   \overset{\mathrm{DLA}}{=}\mathcal H.
 \end{aligned}
 \label{eq:rung-eigenvalue}
\end{equation}
Differentiating the upper integration limit produces a boundary term without a transverse-momentum integral. It therefore lacks one logarithm and is subleading at DLA. Since we have made no restriction on the number of rungs in $\mathcal H_{\rm sub}$, the same argument applies at every order: we can obtain the SGP contribution by transverse differentiation without changing the momentum dependence of the unpolarized ladder kernel.

\begin{figure}[!htbp]
 \centering
 \includegraphics[width=\columnwidth]{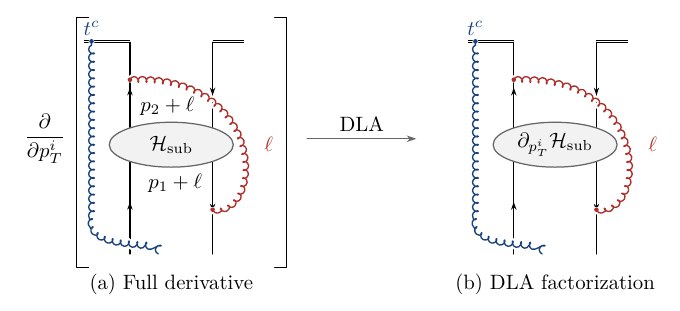}
 \caption{Nonladder factorization. (a) Transverse differentiation of the full soft-gluon contribution. (b) Equivalent form at DLA, with differentiation restricted to the harder subamplitude.}
 \label{fig:sgp-nonladder}
\end{figure}

We next consider nonladder radiation and show that we can factor out the softest gluon before differentiating with respect to the coherent-gluon transverse momentum. For nonladder diagrams, a leading double-logarithmic contribution arises when the exchanged gluon has the smallest transverse momentum in the diagram. In this region, we obtain one logarithm from the longitudinal integration and another from the transverse integration. We isolate this gluon and write its momentum in light-front components as $\ell=(\ell^+,\ell^-,\bm\ell_T)$, with $s=2p_1^+p_2^-$. Here $p_1$ and $p_2$ are lightlike momenta along the plus and minus directions, respectively.

After summing over all soft-gluon attachments, we use a Ward identity to express the result in terms of eikonal couplings to the external legs of the harder subamplitude~\cite{Ermolaev:1995fx,Bartels:1995iu}. We denote the harder block by $\mathcal H_{\rm sub}(\bm k_T,\bm p_T,\bm\ell_T)$.

To examine the effect of transverse differentiation, we consider the attachment shown in Fig.~\ref{fig:sgp-nonladder}, assigning the transverse momentum $\bm p_T$ to the external leg along $p_1$. To first order in $\bm p_T$, the kinematic part of its eikonal factor is
\begin{equation}
 \mathcal S(\bm p_T,\ell)
 \simeq\frac{s/2}
 {[p_1\cdot\ell-\bm p_T\cdot\bm\ell_T+i0]
  [p_2\cdot\ell+i0]}.
 \label{eq:soft-eikonal-factor}
\end{equation}
Expanding at fixed $\ell$, we find
\begin{equation}
 \mathcal S(\bm p_T,\ell)
 \simeq \mathcal S(0,\ell)
 \left[1+\frac{\bm p_T\cdot\bm\ell_T}{p_1\cdot\ell+i0}\right].
 \label{eq:soft-eikonal-expansion}
\end{equation}
At the soft-gluon propagator pole, $\ell^-=\ell_T^2/(2\ell^+)$, we therefore obtain
\begin{equation}
 \left.\frac{\partial \mathcal S(\bm p_T,\ell)}{\partial p_T^i}
 \right|_{\bm p_T=0}
 =\mathcal S(0,\ell)\frac{2\ell^+\ell_T^i}{p_1^+\ell_T^2}.
 \label{eq:soft-kernel-transverse-derivative}
\end{equation}
Differentiating the soft factor therefore supplies an extra factor of $\ell^+/p_1^+$. At DLA, we consequently need to differentiate only the harder subamplitude. We can also neglect the soft transverse recoil in this derivative: the recoil corrections are suppressed by the ratio of soft to hard transverse scales and do not produce the transverse logarithm.

Using these DLA approximations, we obtain the factorized expression for the attachment in Fig.~\ref{fig:sgp-nonladder}, leaving the common coupling and color factors implicit:
\begin{align}
 &\left.\frac{\partial}{\partial p_T^i}
 \int\frac{\mathrm d^4\ell}{(2\pi)^4}\,
 \frac{-i}{\ell^2+i0}\,
 \mathcal S(\bm p_T,\ell)\mathcal H_{\rm sub}(\bm p_T,\bm\ell_T)
 \right|_{\bm p_T=0}
 \label{eq:soft-derivative-recursion}\\
 &\quad\overset{\mathrm{DLA}}{=}
 \int\frac{\mathrm d^4\ell}{(2\pi)^4}\,
 \frac{-i}{\ell^2+i0}\,\mathcal S(0,\ell)
 \left.\frac{\partial\mathcal H_{\rm sub}(\bm p_T,0)}{\partial p_T^i}
 \right|_{\bm p_T=0} .\notag
\end{align}
This factorization also holds for attachments of the softest gluon to the other external legs of the harder subamplitude. We repeat this procedure within the harder subamplitude, successively factoring out the nonladder gluons in order of increasing transverse momentum. At each step, the transverse derivative acts only on the remaining harder subamplitude. When the softest remaining emission belongs to the ladder, we evaluate the derivative explicitly using Eq.~\eqref{eq:rung-eigenvalue}; if only the Born hard part remains, we differentiate it directly. Throughout this iteration, we keep the color matrices in their original order.

\emph{Infrared evolution equations.}---Equations~\eqref{eq:rung-eigenvalue} and \eqref{eq:soft-derivative-recursion} allow us to use the same auxiliary-amplitude formulation as in unpolarized infrared evolution. The recursive construction remains unchanged; for the QS correlation, we need to determine the appropriate color factors and boundary conditions.

\begin{figure}[t]
 \centering
 \includegraphics[width=\columnwidth]{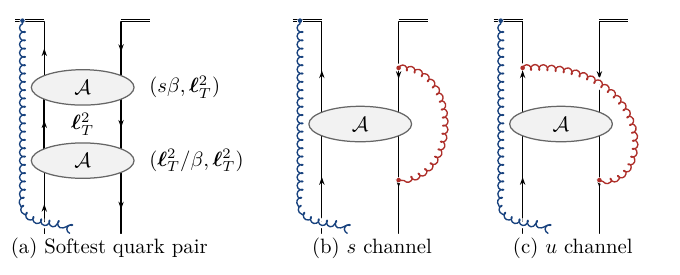}
 \caption{Factorization at the smallest transverse scale generates the coupled high-energy evolution~\cite{Bartels:1995iu}. (a) A softest quark pair separates two harder subamplitudes with cutoff $\bm\ell_T^2$. (b),(c) Soft-gluon attachments in the $s$ and $u$ channels generate color and signature mixing.}
 \label{fig:iree-softest-lines}
\end{figure}

We represent the harder blocks by near-forward amplitudes $\mathcal A_r^{(\pm)}$, with color-singlet ($r=1$) or octet ($r=8$) quark--antiquark exchange and signature $\pm$ under $s\leftrightarrow u$. To construct their evolution, we isolate the intermediate state with the smallest transverse momentum. For a quark--antiquark pair, we factorize the diagram into two harder subamplitudes connected by the pair [Fig.~\ref{fig:iree-softest-lines}(a)]. For a nonladder gluon, we factor out its eikonal coupling to the external legs of the remaining harder subamplitude and sum the attachments in the $s$ and $u$ channels [Figs.~\ref{fig:iree-softest-lines}(b),(c)].

To resum the energy logarithms, we introduce Mellin partial waves $h_r^{(\pm)}(\omega)$ with signature factors $\eta^\pm(\omega)=(e^{-i\pi\omega}\pm1)/2$:
\begin{equation}
 \mathcal A_r^{(\pm)}(s,\mu^2)
 =8\pi^2\int_{c-i\infty}^{c+i\infty}\frac{\mathrm d\omega}{2\pi i}
 \left(\frac{s}{\mu^2}\right)^\omega\eta^\pm(\omega)h_r^{(\pm)}(\omega),
 \label{eq:mellin-rep}
\end{equation}
where the contour lies to the right of all singularities. The corresponding Mellin moments are defined for $x>0$ as $T_F^{q,\bar q}(\omega;Q^2)=\int_0^1\mathrm dx\,x^{\omega-1}T_F^{q,\bar q}(x,x;Q^2)$.

We now differentiate the auxiliary amplitudes with respect to the infrared cutoff to derive their evolution equations. For the quark--antiquark contribution in Fig.~\ref{fig:iree-softest-lines}(a), we obtain terms quadratic in the partial waves from the product of the two harder subamplitudes. After summing the soft-gluon attachments in Figs.~\ref{fig:iree-softest-lines}(b),(c), we also obtain $\omega$-derivatives and mixing between color and signature channels. Following the standard IREE procedure~\cite{Ermolaev:1995fx,Bartels:1995iu}, we obtain the coupled equations for $h_1^{(+)}$, $h_8^{(+)}$, and $h_8^{(-)}$:
\begin{align}
 \omega h_1^{(+)}&=\frac{2\alpha_s}{3\pi}+(h_1^{(+)})^2,
 \label{eq:iree-h1p}\\
 \omega h_8^{(+)}&=-\frac{\alpha_s}{12\pi}
 +\frac{3\alpha_s}{2\pi}\partial_\omega h_8^{(+)}+(h_8^{(+)})^2,
 \label{eq:iree-h8p}\\
 \omega h_8^{(-)}&=-\frac{\alpha_s}{12\pi}+(h_8^{(-)})^2
 +\frac{3\alpha_s}{2\pi}\left(\frac{\partial h_8^{(-)}}{\partial\omega}+\frac{h_8^{(-)}}{\omega}\right)
 \notag\\
 &\quad-\frac{\alpha_s}{6\pi\omega}\left(2h_1^{(+)}+5h_8^{(+)}\right).
 \label{eq:iree-h8m}
\end{align}
The Born terms in Eqs.~\eqref{eq:iree-h1p}--\eqref{eq:iree-h8m} are fixed by matching to the one-loop QS evolution kernel~\cite{Braun:2009mi}. Taking the small-$z$ limit, with $z=x/\xi$, and omitting terms localized at $z=1$, we obtain
\begin{equation}
 \frac{\partial T_F^q(x,x;Q^2)}{\partial\ln Q^2}
 \simeq-\frac{\alpha_s}{4\pi N_c}\int_x^1\frac{\mathrm d\xi}{\xi}
 T_F^q(\xi,\xi;Q^2)+\cdots.
 \label{eq:qs-one-loop-small-x}
\end{equation}
Here $\xi$ is the parent quark momentum fraction. The ellipsis denotes contributions omitted at DLA, including the off-diagonal hard-gluon-pole terms.

The color factor $-1/(2N_c)$ arises from the coherent-gluon insertion, which carries a color matrix $t^c$. For the singlet auxiliary amplitude, the quark--antiquark pair is projected onto $\mathbf1$ instead, and a single rung gives $\sum_a t^a\mathbf1t^a=C_F\mathbf1$. The kinematic factor is unchanged, so the singlet Born term follows by replacing $-1/(2N_c)$ with $C_F$. We obtain $2\alpha_s/(3\pi)$ in Eq.~\eqref{eq:iree-h1p} and $-\alpha_s/(12\pi)$ in both Eqs.~\eqref{eq:iree-h8p} and \eqref{eq:iree-h8m}.

To determine the signature under $s\leftrightarrow u$ of the color-octet amplitude relevant to QS evolution, we use the crossing relation $T_F^q(-x,-x)=T_F^{\bar q}(x,x)$~\cite{Koike:2009ge}. The quark--antiquark sum is therefore even under $x\to -x$. With our signature convention, the positive- and negative-$x$ contributions combine to give:
\begin{equation}
 \begin{aligned}
 &\frac12\int_{-1}^{1}\mathrm dx\,x^{\omega-1}
 \bigl[T_F^q(x,x)+T_F^q(-x,-x)\bigr]\\
 &\quad=-\eta^-(\omega)
 \bigl[T_F^q(\omega)+T_F^{\bar q}(\omega)\bigr].
 \end{aligned}
 \label{eq:qs-signature-mellin}
\end{equation}
Comparing with Eq.~\eqref{eq:mellin-rep}, we find that the QS correlation couples to the negative-signature partial wave. Consequently, $h_8^{(-)}(\omega)$ acts as the anomalous dimension for the Mellin moment of the flavor-nonsinglet QS function, yielding
\begin{equation}
 T_{F,\mathrm{NS}}(\omega;Q^2)
 =e^{\ln(Q^2/\mu^2)h_8^{(-)}(\omega)}
 T_{F,\mathrm{NS}}(\omega;\mu^2).
 \label{eq:qs-moment-evolution}
\end{equation}

With the octet inputs evaluated at Born level, we obtain the small-$x$ asymptotics:
\begin{equation}
 \begin{aligned}
 &T_{F,\mathrm{NS}}(x,x;Q^2)
 \underset{x\to0}{\propto}\\
 &\quad\ln\frac{Q^2}{\mu^2}
 \left(\frac{Q^2}{\mu^2}\right)^{-0.065\sqrt{3\alpha_s/(2\pi)}}
 \frac{x^{-\sqrt{8\alpha_s/(3\pi)}}}{[\ln(1/x)]^{5/2}}.
 \end{aligned}
 \label{eq:qs-asymptotics}
\end{equation}

For comparison, the corresponding unpolarized nonsinglet distribution behaves as $f_{1,\mathrm{NS}}\propto x^{-\sqrt{8\alpha_s/(3\pi)}}[\ln(1/x)]^{-3/2}$ under the same assumptions~\cite{Ermolaev:1995fx}. At high energy, evolution in the positive-signature singlet channel produces faster growth than pure octet evolution. The ordinary unpolarized quark distribution function couples directly to the singlet channel. For the QS function, the dominant contribution instead comes from a harder singlet block coupled to a soft nonladder gluon. In the factorized description, this gluon connects external legs of the block and carries the octet color, allowing the harder quark--antiquark pair to evolve as a singlet while the full contribution matches the negative-signature octet channel selected by QS.

To make this physical picture explicit, we focus on the parts of the partial waves that determine the leading small-$x$ behavior and simplify Eq.~\eqref{eq:iree-h8m} to
\begin{equation}
 \frac{\partial h_8^{(-)}}{\partial\omega}
 \simeq\frac{2}{9\omega}\,h_1^{(+)}.
 \label{eq:qs-leading-balance}
\end{equation}
Under inverse Mellin transformation, the derivative supplies a factor of $-\ln(1/x)$. The QS function therefore inherits the singlet's leading small-$x$ power, but acquires an additional $1/\ln(1/x)$ suppression from the octet virtual corrections.

We can apply the same method directly to the chiral-odd quark--gluon correlator $T_F^{(\sigma)}$ associated with the Boer--Mulders function~\cite{Boer:1997nt,Boer:2003cm}. The corresponding twist-three hard parts have the same pole structure as in the QS case, so transverse differentiation does not disrupt recursive factorization.

We take the flavor combination $T_{F,\mathrm{NS}}^{(\sigma)}=T_F^{(\sigma),u}+T_F^{(\sigma),\bar u}-T_F^{(\sigma),d}-T_F^{(\sigma),\bar d}$. The real-emission kernel for soft-gluon-pole evolution contains $2z/(1-z)$~\cite{Zhou:2008fb,Zhou:2008mz,Zhou:2009rp}, instead of $(1+z^2)/(1-z)$ in the QS case. Factoring out $z$ leaves a reduced chiral-odd amplitude with the same IREE recursion and crossing symmetry as in the QS case, but with twice the Born input. To restore the factor $z$ in Mellin space, we describe the evolution using the shifted partial wave $h_{8,\sigma}^{(-)}(\omega+1)$.

Solving the infrared evolution equation, we obtain the small-$x$ asymptotic behavior
\begin{equation}
 \begin{aligned}
 &T_{F,\mathrm{NS}}^{(\sigma)}(x,x;Q^2)
 \underset{x\to0}{\propto}\\
 &\quad\ln\frac{Q^2}{\mu^2}
 \left(\frac{Q^2}{\mu^2}\right)^{-0.080\sqrt{3\alpha_s/(2\pi)}}
 \frac{x^{1-\sqrt{16\alpha_s/(3\pi)}}}{[\ln(1/x)]^{5/2}}.
 \end{aligned}
 \label{eq:bm-asymptotics}
\end{equation}
Compared with the transversity result $h_{1,\mathrm{NS}}(x;Q^2)\propto x^{1-\sqrt{16\alpha_s/(3\pi)}}[\ln(1/x)]^{-3/2}$ of Ref.~\cite{Kirschner:1996jj}, we again find an additional $1/\ln(1/x)$ suppression.

\emph{Conclusions.}---We have established, for the first time, the high-energy asymptotics of quark--gluon--quark correlators. These results establish a dynamical connection between single-parton densities and multiparton quantum interference in the high-energy limit. The coherent-gluon insertion is essential: the associated color structure governs how quark--gluon interference evolves at high energy. Beyond the T-odd correlations studied here, this approach offers a route to the small-$x$ evolution of the twist-three correlations $\tilde g(x)$ and $\tilde h(x)$ associated with T-even TMDs~\cite{Zhou:2009jm}. More broadly, our results provide new insight into the proton's multiparton structure, supporting the spin program at the EIC.

\noindent{\it Acknowledgments.} This work was supported by the National
Natural Science Foundation of China under Grant Nos.~12175118, 12321005,
and 12575091 (J.Z.).

\nocite{*}
\bibliographystyle{apsrev}
\bibliography{ref}

\end{document}